\documentclass[prl,twocolumn,superscriptaddress,showpacs,floatfix]{revtex4}
\usepackage{graphics}
\usepackage{graphicx}
\begin{document}
\newcommand{\RR}{\mathrm{\mathbf{R}}}
\newcommand{\rr}{\mathrm{\mathbf{r}}}
\newcommand{\defin}{\stackrel{def}{=}}
\title{Magnetic field-assisted  manipulation and entanglement of Si
spin qubits}
\author{M.J. Calder\'on}
\affiliation{Condensed Matter Theory Center, Department of Physics,
University of Maryland, College Park, MD 20742-4111}
\author{Belita Koiller}
\affiliation{Condensed Matter Theory Center, Department of Physics,
University of Maryland, College Park, MD 20742-4111}
\affiliation{Instituto de F\'{\i}sica, Universidade Federal do Rio de
Janeiro, Caixa Postal 68528, 21941-972 Rio de Janeiro, Brazil}
\author{S. Das Sarma}
\affiliation{Condensed Matter Theory Center, Department of Physics,
University of Maryland, College Park, MD 20742-4111}
\date{\today}

\begin{abstract}
Architectures of donor-electron based qubits in silicon
near an oxide interface are considered theoretically.
We find that the precondition for reliable logic and read-out
operations, namely the individual identification of each donor-bound electron
near the interface, may be accomplished by fine-tuning electric and magnetic
fields, both  applied perpendicularly to the interface.
We argue that such magnetic fields may also be valuable in controlling
two-qubit entanglement via donor electron pairs near the interface.
\end{abstract}

\maketitle

\begin{figure}
\resizebox{70mm}{!}{\includegraphics{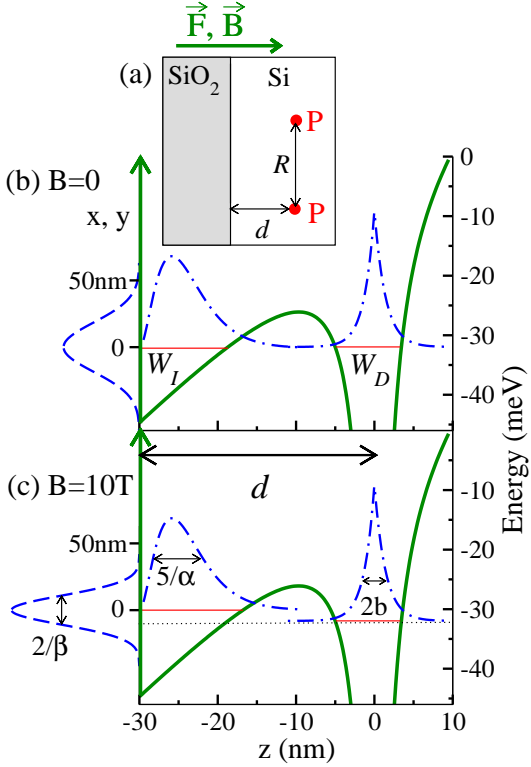}}
\caption{\label{fig:diagrams}(Color online)
(a) Schematic  configuration of donors in Si near the interface with an oxide
 barrier under applied electric ${\bf F}$ and magnetic ${\bf B}$
 uniform fields. 
In (b) and (c), the thick lines show the electronic
confining  double-well potential: The interface well is labeled $W_I$
and the donor well, $W_D$.  The dashed lines give the decoupled
ground-eigenfunctions in the wells, and the thin horizontal lines
indicate the  expectation value of the energy in each well.
Results for $d=30$\,nm and  $F = 13.5$\,kV/cm\,$\approx F_c(d)$,
are given in (b) for $B=0$ and in (c) for $B=10$\,T.
Parameters defining typical wavefunction widths (see text) are
indicated in (c). Magnetic field effects tend to be stronger for
 larger $d$, as the interface state is less affected by the strongly
 confining Coulomb potential at $W_D$.
}
\end{figure}
Spin qubits
in semiconductors (e.g.~GaAs, Si) are among the most promising
physical systems for the eventual fabrication of a
working quantum computer (QC). There are two compelling reasons for this
perceived importance of semiconductor spin qubits: (i) electron
spin has very long coherence times, making quantum error correction
schemes feasible as a matter of principle; (ii) semiconductor
structures provide inherently scalable solid state architectures, as
exemplified by the astonishing success of the microelectronics
technology in increasing the speed and efficiency of logic and memory
operations over the last fifty years (i.e. `Moore's Law'). 
These advantages of semiconductor quantum
computation apply much more to Si than to GaAs, because the electron
spin coherence time can be increased indefinitely (up to 100 ms or
even longer in the bulk) in Si through isotopic
purification~\cite{sousa03,witzel05,abe04} whereas in GaAs the electron
spin coherence time is restricted only to about 10 $\mu
s$~\cite{sousa03,witzel05,petta05}. It is thus quite ironic that there
has been much 
more substantial experimental progress~\cite{petta05} in
electron spin and charge
qubit manipulation in the III-V semiconductor quantum dot systems~\cite{Exch}
than in the Si:P 
Kane computer architecture~\cite{Kane}. In addition to the long
coherence time, the Si:P architecture
has the highly desirable property of 
microscopically identical qubits which are scalable using the Si
microelectronic technology.
The main reason for the slow
experimental progress in the Kane
architecture is the singular lack of qubit-specific quantum control 
over an electron which is localized around
a substitutional P atom in the bulk in a relatively unknown location. This control has
turned out to be an 
impossibly difficult experimental task in spite of impressive
developments in materials fabrication and growth in the Si:P
architecture using both the `top-down' and the `bottom-up'
techniques~\cite{schenkel03}. 
It is becoming manifestly clear that new ideas are
needed in developing quantum control over single qubits in the Si:P
QC architecture.

In this Letter we suggest such a new idea, establishing convincingly
that the use of a magnetic field, along with an electric field, would
enable precise identification, manipulation and entanglement of donor qubits in the
Si:P quantum computer architecture by allowing control over the
spatial location of the electron as it is pulled from its shallow
hydrogenic donor state to the Si/SiO$_2$ interface by an electric
field. Additionally, the magnetic field could be used to control 
the spatial overlap of the electronic wavefunctions in the 2D plane
parallel to the interface (i.e. similar to a MOSFET geometry), thus
enabling external manipulation of the inter-qubit entanglement through
the magnetic field tuning of the exchange coupling between neighboring
spin qubits~\cite{Exch}.
 

Fig.~\ref{fig:diagrams} highlights the basic physical effects explored
in our theoretical study:
Substitutional P donors in Si, separated by $R$, are a distance
$d$ from an ideally flat Si/SiO$_2$ $(001)$ interface. Uniform
electric ${\bf F}$ and magnetic ${\bf B}$ fields are both applied along $z$.
When a single P atom is considered  ($R\to \infty$ limit), and in the
absence of external fields, the active electron is bound to the donor
potential well ($W_D$) forming a hydrogenic atom. If a uniform
electric field is also present, an additional well is formed
which tends to draw the donor electron toward the interface.
The interface well ($W_I$) is triangular-shaped along $z$
while in the $xy$ 2D plane the confinement is still provided by the `distant' donor
Coulomb attraction. The ground-state wavefunctions for each well are
given in Fig.~\ref{fig:diagrams}(b). This corresponds to a particular
value of the applied field, the critical field which we denote by
$F=F_c(d)$, such that the expectation values of the energy for the
$W_I$ and $W_D$ ground-eigenstates are degenerate.
Fig.~\ref{fig:diagrams}(c) shows how the system in (b)
changes due to a magnetic field applied along $z$. The interface
ground state wavefunction undergoes the usual `shrinking'
perpendicular to ${\bf B}$, and the energy at $W_I$ is raised,
while no significant changes occur at $W_D$.
As a consequence, we find that by properly tuning ${\bf B}$ it is
possible to control the electron location {\it along the applied
 magnetic field direction}, partially or completely reversing the
effect of the electric field.

We base our quantitative description of this problem
on the single-valley effective-mass approximation,
leading to the model-Hamiltonian~\cite{macmillen84,calderonPRL06}
\begin{equation}
H = T + e F z -{{e^2}\over{\epsilon_1 r}}+{{e^2
Q}\over{\sqrt{\rho^2+(z+2d)^2}}}-{{e^2 Q}\over{4(z+d)}}~.
\label{eq:h}
\end{equation}
The magnetic field vector potential,
${\bf A} = B \left(y,-x,0 \right)/2$,
is included in the kinetic energy term,
 $T=\sum_{\eta=x,y,z} \hbar^2/(2 m_\eta) \left(i
\partial/\partial  \eta+e
A_\eta/(\hbar c) \right)^2 $, where the effective masses
 $m_x=m_y=m_\perp = 0.191 \,m$ and $m_z=m_\|=0.916\,m$, account
for the Si conduction band valley's anisotropy.
The electric field defines the second term in $H$, the
third and fourth terms describe the donor and its image charge
potentials, while the last term is the electron image potential.
The fourth term involves the lateral coordinate $\vec\rho=(x,y)$.
The image-related terms are proportional to
$Q={{(\epsilon_2-\epsilon_1)}/{[\epsilon_1 (\epsilon_2+\epsilon_1)]}}$,
where $\epsilon_1=11.4$ and  $\epsilon_2=3.8$ are the Si and SiO$_2$
static dielectric constants.
It is convenient to rewrite the kinetic energy term as
\begin{eqnarray}
T= T_0 + {{1}\over{8}} \omega_c^2\rho^2+i \omega_c\left(y
{{\partial}\over{\partial  x}} -x
{{\partial}\over{\partial  y}} \right)~,
\end{eqnarray}
where $T_0$ is the kinetic energy for $B=0$ and
$\omega_c=e B /(m_\perp c)$.
The second term allows interpreting the effect of the
magnetic field as providing an additional parabolic potential in the
$xy$ plane, which increases the kinetic energy and enhances the
lateral confinement of the electron wave-function.
The last term in $T$ gives zero or negligible contribution for the
donor electron low-energy states being investigated, and will be
neglected here.

We solve the double-well problem described by the model Hamiltonian
in Eq.~(\ref{eq:h}) in the basis of the uncoupled solutions to the
individual wells $W_I$ and $W_D$, obtained variationally from the {\it
 ansatz}~\cite{calderonPRL06}:
\begin{eqnarray}
\Psi_I(\rho, z)&=&f_\alpha(z)\times g_\beta(\rho)\\
&=&{{\alpha^{5/2}}\over{\sqrt{24}}} (z+d)^{2}\, e^{-{\alpha
(z+d)}/{2}}\times{{\beta}\over{\sqrt{\pi}}}\, e^{{-\beta^2
\rho^2}/{2}}
\label{eq:psia}\nonumber \\
\Psi_D (\rho, z)&\propto&(z+d)e^{-\sqrt{\rho^2 / a^2+z^2/b^2}}~,
\label{eq:psib}
\end{eqnarray}
where $\alpha,~\beta,~a$ and $b$ are the variational parameters.
The $(z+d)$ factors guarantee that the wavefunctions are zero at the
interface $(z=-d)$, as we assume that the insulator provides an
infinite barrier potential.
For $d> 6$ nm and $B=0$, we find that $a$ and $b$ coincide with the
Kohn-Luttinger variational Bohr radii for the isolated impurity
($d\rightarrow \infty$), where $a=2.365$ nm and $b=1.36$ nm.

As an illustration of the uncoupled wells variational solutions, we
present in Fig.~\ref{fig:diagrams} the calculated wavefunctions for
$d=30$nm. The dash-dot lines correspond to
$f_\alpha(z)$ and $\Psi_D (0, z)$, while the dashed lines on the left
represent $g_\beta(\rho)$ for $\vec\rho=(x,0)$ [or equivalently $(0,y)$].
It is clear from this figure that the four variational parameters define
relevant length scales involved in the problem.
In the presence of a magnetic field, there is an additional length scale
(the magnetic length) given by $\lambda_B=\sqrt{\hbar c/(eB)}$ which
defines the typical lateral confinement produced
by the magnetic field alone, regardless of other potentials in the
problem. The magnetic field effect is significant
only for values of $B \gtrsim B_c$ such that $\lambda_B \lesssim c_\rho$
where $c_\rho$ is the lateral confinement length in the absence of the
field.  For the donor potential well, $c_\rho \approx a=2.365$ nm,
so that the field required to appreciable affect the $W_D$-ground
state eigenenergy or wavefunction is $B^D_c \sim 120$ T!
On the other hand, for the  $W_I$ well, the confinement length at the interface
for $B=0$ is $1/\beta \sim 18.5$ nm so that a much smaller
$B^I_c \sim 2$ T is sufficient to affect the interface ground state.
The magnetic field strengths we consider here (up to $10$ T)
are not large enough to affect $a$ or $b$, but important effects are
obtained in lateral confinement at the interface as shown in
Fig.~\ref{fig:diagrams} by comparison of $B=0$ in (b) with $B=10$ T in
(c).

\begin{figure}
\resizebox{74mm}{!}{\includegraphics{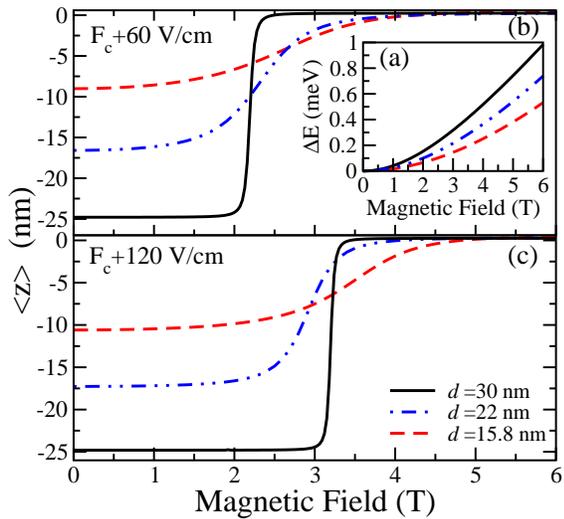}}
\caption{\label{fig:zav}(Color online) 
(a) Shift in the energy ${\cal E}_I$, associated with the interface state,
versus magnetic field for three values of the donor distance to the
barrier ($d$). The shift in energy of the donor state (${\cal E}_D$) is
negligible compared to these values and is not shown. (b) Expectation
value of the $z$-position for the electronic ground state versus magnetic
field. For each $d$ the electric field is tuned to a value just above the
critical field: $F = F_c+ 60$ V/cm. The electron starts at a position
close to the interface and moves parallel to the magnetic field and
against the electric field ending up at the donor well. The values of the
critical electric field at each distance are $F_c (15.8$ nm$)=27.44$
kV/cm, $F_c (22$ nm$)=18.85$ kV/cm, $F_c (30$ nm$)=13.44$ kV/cm. (c) Same
as (b) for $F = F_c+ 120$ V/cm.
}
\end{figure}
The expectation value of the energy for each well ground state,
${\cal E}_j =\langle \Psi_j|H|\Psi_j \rangle_{(j=I,D)}$,
is given by the horizontal lines in  Fig.~\ref{fig:diagrams}. The
critical field condition $F=F_c$ corresponds to $ {\cal E}_I={\cal
E}_D$ and is illustrated in Fig.~\ref{fig:diagrams}(b).
In Fig.~\ref{fig:diagrams}(c) we also indicate
the value of the $B=0$ energies  by a thin dotted line, and we note that
under a 10 T magnetic field ${\cal E}_I$ is
raised by 2 meV,  while ${\cal E}_D$ undergoes a comparatively
negligible shift (by 0.13 meV). Fig.~\ref{fig:zav}(a) shows the energy
shifts, $\Delta E (B) ={\cal E}_I (B)-{\cal E}_I (0)$, for three
particular values of $d$: As expected, the effect of the magnetic
field is stronger for larger $d$.

Consequences of this shift in energy 
are shown in Figs.~\ref{fig:zav}(b) and (c) where we give
the expectation value of the $z$-coordinate of the electronic ground
state $\Psi_0= C_I \Psi_I+C_D \Psi_D$.  We start with an
electric field slightly above $F_c(d)$, hence, $C_I \approx 1$ and
$C_D \approx 0$.
Under an increasing magnetic field, the energy shift
at the interface
eventually detunes ${\cal E}_D$ and
${\cal E}_I$ in
such a way that $C_I \approx 0$ and
$C_D \approx 1$, i.e. the electron is taken back to the donor moving
parallel to the magnetic field ${\bf B}$ and against the electric
field ${\bf F}$. How big 
a magnetic field is needed for this process to be completed depends on
$d$ and on the value of the static electric field. 
For those donors further
from the interface, the passage of the electronic ground state 
from the $W_I$ to $W_D$
occurs more abruptly as a function of $B$.
The effect of the value of the electric field is also
relevant, as illustrated
in Fig.~\ref{fig:zav}(b) and (c). In
Fig.~\ref{fig:zav}(b), $F=F_c+60$ V/cm and the passage occurs 
around $B \sim 2.2$ T, while in  Fig.~\ref{fig:zav}(c), $F=F_c+120$ V/cm and
the needed field is $B \sim 3.2$ T. Therefore, fine tuning of
the electric field is required to observe this phenomenon at
reasonably small magnetic fields. We propose to use this result as a
means of differentiating donor electrons from other charges that may
be detected~\cite{Kane_MRS} in the architecture shown in
Fig.~\ref{fig:diagrams}(a). 
Tunneling times are not significantly affected upon
 magnetic fields considered here  and remain a function of
$d$ alone~\cite{calderonPRL06}. 

\begin{figure}
\begin{center}
\resizebox{78mm}{!}{\includegraphics{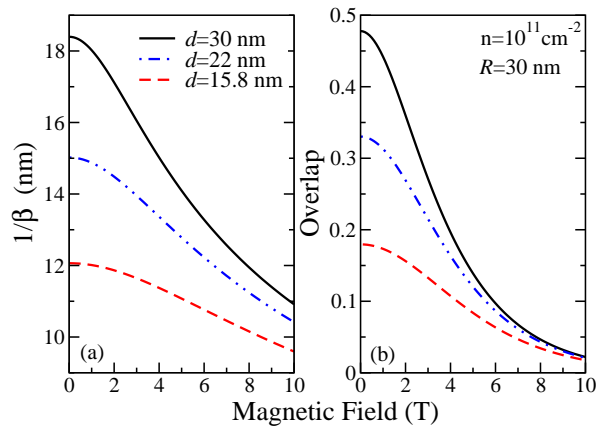}}
\caption{\label{fig:beta-S}(Color online) 
(a) Reduction of the lateral confinement length of the interface state,
$1/\beta$, when a perpendicular magnetic field is applied. The effect is
stronger for donors that are further away from the interface. (b) Overlap
$S$ versus magnetic field for inter-donor separation $R=30 nm$ (which
corresponds to a planar density $n=10^{11}$cm$^{-2}$).
}
\end{center}
\end{figure}
\begin{figure}
\begin{center}
\resizebox{76mm}{!}{\includegraphics{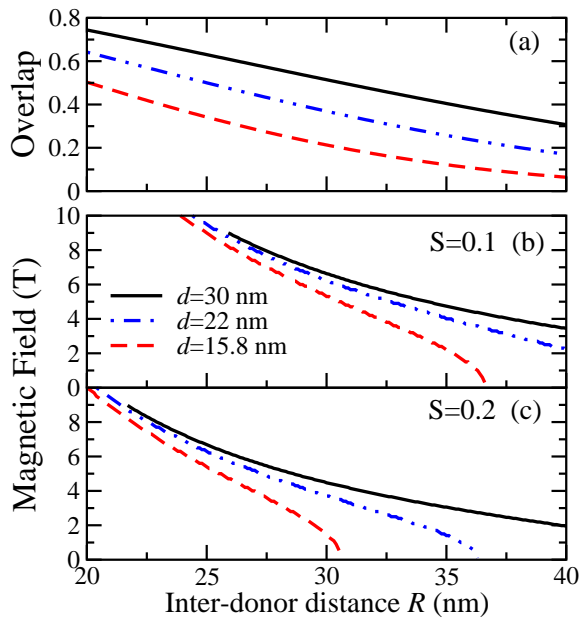}}
\caption{\label{fig:S}(Color online) (a) Overlap versus interdonor
distance. (b) and (c) show the magnetic field required to reduce the
overlap to $0.1$ and $0.2$ respectively. $R=20$ nm corresponds to a
planar density $n=2.5\times10^{11}$ cm$^{-2}$ and $R=40$ nm to $n=6.25\times10^{10}$ cm$^{-2}$.
}
\end{center}
\end{figure}
The other significant effect of the magnetic field on the interface states is
the transverse shrinkage of the wave function. Fig.~\ref{fig:beta-S}(a) shows the lateral confinement length $1/\beta$ as a
function of the magnetic field. For $d=15.8$ nm, the wave-function width
is
reduced by a  factor of $0.8$ when a $10$ T magnetic field is
applied, while for $d=30$ nm a stronger reduction factor of $0.6$ is obtained. 
The significance of this enhanced
confinement may be accessed through the overlap between interface electrons 
coming
from neighboring donors separated a distance $R$. The overlap
quantifies whether we can consider these electrons as separate
entities or they will form a 2-dimensional electron gas (2DEG). 
One obtains $S=\langle g_{\beta}(\vec
\rho)|g_{\beta}(\vec \rho +\vec R) \rangle=\exp(-\beta^2 R^2/4)
\exp[-R^2/(16\lambda_B^4 \beta^2)]$ where we note that, due to gauge
invariance,  $g_{\beta}(x + R,
y)=\pi^{-1}\beta  \exp\{-\beta^2[(x+R)^2+y^2]/2\} \exp{\{i y R/(4
\lambda_B^2)\}}$. The results
for the particular inter-donor distance $R=30$ nm, which corresponds to a
planar density $n \sim 10^{11}$ cm$^{-2}$, are shown in
Fig.~\ref{fig:beta-S}(b). We estimate that an overlap $S \lesssim 0.1$
would guarantee that the electrons do not form a 2DEG.  
Fig.~\ref{fig:S}(a) presents
the overlap versus inter-donor distance for $B=0$. The calculated $S$ is generally
very large for the experimentally reasonable distances $d$ and $R$ considered here. 
For the donors further from the interface ($d=30$ nm), the
condition $S<0.1$ requires either the application of a magnetic field (around
$6$ T for $R=30$ nm) or a smaller planar density $n
\lesssim 3\times 10^{10}$ cm$^{-2}$ (which corresponds to $R \gtrsim
55 nm$). 
The magnetic fields required to get a reduction of the overlap to $S=0.1$
and $0.2$ are shown in Fig.~\ref{fig:S}(b) and (c), respectively, as a
function of the inter-donor distance. For the donors closer to the
interface ($d=15.8$ nm), the effect of the magnetic field is not as
dramatic: The lateral confinement provided by the donor potential 
is strong enough to give
$S=0.1$ for $B=0$ and $R \sim 37$ nm ($n \sim 7.4\times 10^{10}$
cm$^{-2}$). 
Note that in the case of neighboring donors placed at different
distances from the interface $d$, $S(R)$ would present
oscillations due to valley interference effects~\cite{KHD2}. 
 
For qubits defined at the donor sites it is certainly more reliable to
perform operations involving two-qubit entanglement at the more
directly accessible interface region. In this case, depending on $R$
and $d$, the confining potential provided by the donors may need to be
complemented by additional surface gates~\cite{footgates} and/or static magnetic
fields. Time-dependent magnetic fields can be invoked in switching the
exchange gates for two-qubit operations~\cite{Exch}.

In summary, we demonstrate here that, in addition to the control of
donor charges by electric fields, as usual in conventional
semiconductor devices, relatively moderate magnetic fields
(up to 10 T) may provide relevant information and
manipulation capabilities as the building blocks of
donor-based QC architectures in Si are being developed.
Uniform magnetic fields may displace donor-bound electrons from the
interface to the donor nucleus region, as illustrated in Fig.~\ref{fig:zav}. This
effect, which could be monitored via surface charge detectors,
allows  differentiating donor electrons from spurious surface or oxide-region
bound electrons, which would not exhibit this type of behavior.
For electrons originating from neighboring donors and drawn to the
surface by an electric field, the magnetic field lateral confinement
effect may provide isolation between electrons (see
Fig.~\ref{fig:beta-S}), allowing smaller 
interdonor distances to be exploited (see Fig.~\ref{fig:S}), as well
as additional 
control in two qubits operations via surface exchange gates similar to
the spin manipulation in 
quantum dots in GaAs.  Defects such as dangling bonds at the
Si/SiO$_2$ interface are 
important sources of spin decoherence, thus spin coherence times near
the interface are expected to be significantly reduced as compared to
the bulk. Performing two-qubit (spin-spin) entanglement near the SiO$_2$
interface, as suggested here, requires extremely careful interface
optimization. We believe that the use of an external magnetic
field along with the FET geometry near a Si-SiO$_2$ interface should
allow in the near future 
significant experimental progress in the currently stalled, but
potentially important, donor qubit based Kane spin QC
architecture.

This work is supported by LPS and NSA. BK
acknowledges support by CNPq and 
FAPERJ.

\bibliography{mag}

\begin{thebibliography}{20}
\bibitem[{foo()}]{footgates}
\bibinfo{note}{For typical values of $R$ and $d$, the double-well potential
  from the donor pair alone usually provides a relatively shallow barrier,
  resulting in negligibly small values of $J$. Surface gates between the donors
  could raise the barrier and allow closer donor (smaller $R$) geometries,
  leading to values of $J$ compatible with the required gating times ($\hbar/J
  \ll$ $\sim$ ms), similar to the situation in GaAs quantum dot quantum
  computing architectures (Ref.~\onlinecite{petta05})}.
\end{thebibliography}

\end{document}